\documentclass[twocolumn,         
               showkeys,showpacs,            
               preprintnumbers,longbibliography,     
               aps,                 
               prd,          	    
               letterpaper,             
               nofootinbib,         
               tightenlines,        
               floats,floatfix      
               ]{revtex4-1}
\usepackage{dcolumn}   
\usepackage{bm}        
\usepackage{float}
\usepackage{physics}
\usepackage{graphicx,color}
\usepackage{subfigure}
\usepackage{latexsym}
\usepackage{amsmath,amssymb}        
\usepackage[colorlinks=true,linkcolor=blue,citecolor=blue]{hyperref}
\usepackage{mathrsfs}
\usepackage{comment}
\usepackage{soul}
\usepackage{cancel}
\definecolor{purple}{rgb}{0.58,0.0,0.83}
\definecolor{orange}{rgb}{1,0.5,0}
\DeclareSymbolFontAlphabet{\mathrsfs}{rsfs}
\DeclareMathAlphabet{\mathcal}{OMS}{cmsy}{m}{n}

\begin{document}


\title{Stationary solutions of the Schr\"odinger-Poisson-Euler system and their stability}


\author{Iv\'an \'Alvarez-Rios}
\email{ivan.alvarez@umich.mx}
\affiliation{Instituto de F\'{\i}sica y Matem\'{a}ticas, Universidad
              Michoacana de San Nicol\'as de Hidalgo. Edificio C-3, Cd.
              Universitaria, 58040 Morelia, Michoac\'{a}n,
              M\'{e}xico.}

\author{Francisco S. Guzm\'an}
\email{francisco.s.guzman@umich.mx}
\affiliation{Instituto de F\'{\i}sica y Matem\'{a}ticas, Universidad
              Michoacana de San Nicol\'as de Hidalgo. Edificio C-3, Cd.
              Universitaria, 58040 Morelia, Michoac\'{a}n,
              M\'{e}xico.}


\date{\today}


\begin{abstract}
We present the construction of stationary boson-fermion spherically symmetric configurations governed by Newtonian gravity. Bosons are described in the Gross-Pitaevskii regime and fermions are assumed to obey Euler equations for an inviscid fluid with polytropic equation of state. The two components are coupled through the gravitational potential. The families of solutions are parametrized by the central value of the wave function describing the bosons and the central denisty of the fluid. We explore the stability of the solutions using numerical evolutions that solve the time dependent Schr\"odinger-Euler-Poisson system, using the truncation error of the numerical methods as the perturbation. We find that all configurations are stable as long as the polytropic equation of state (EoS) is enforced during the evolution. When the configurations are evolved using the ideal gas EoS they all are unstable that decay into a sort of twin solutions that approach a nearly stationary configuration. We expect these solutions and their evolution serve to test numerical codes that are currently being used in the study of Fuzzy Dark Matter plus baryons.
\end{abstract}


\keywords{dark matter -- Bose condensates -- numerical astrophysics}


\maketitle

\section{Introduction}

In this paper we construct stationary solutions of the Schr\"odinger-Euler-Poisson system with spherical symmetry, intended to describe configurations of bosonic and fermionic matter at the same time. The bosonic part is represented by the parameter order $\Psi$ that obeys the Gross-Pitaevskii equation for a Bose-Einstein Condensate in the mean field approximation and low temperatures \cite{grosspitaevskii}, which is mathematically the same as Schr\"odinger equation. The fermionic part is modeled by Euler equations that describe the dynamics of an inviscid compressible fluid. These two components interact through gravity, whose potential is sourced by the two ingredients simultaneously.
This problem can also be viewed as the weak gravitational field, low energy limit of bosons, low pressure and low velocities of the fermion gas of the General Relativistic (GR) case, where fermions are modeled with a perfect fluid, whereas the bosonic sector results from the Lagrangian for a complex scalar field with conserved charge \cite{HENRIQUES1989}. The GR case is rich in various aspects, for example, equilibrium solutions of the fermion-boson equations have stable and unstable branches, whose configurations oscillate or collapse respectively \cite{HENRIQUES1990,Valdez2013,Brito2016b}, they can collapse to form black holes or, at least in the case of pure bosons can also explode \cite{Guzman2004BS,Guzman2009}. Nevertheless, in our solutions, the bases and motivation lie within different grounds.

The motivation to construct our solutions is its relation to the model of ultralight bosonic dark matter or Fuzzy Dark Matter (FDM), that reproduces the properties of Cold Dark Matter at big scales, whereas it solves some of its problems, like the too-big-to-fail and the cusp-core inconsistencies \citep{Hui:2021tkt,Niemeyer_2020,Lee:2017,Hui:2016,Suarez:2013,Chavanis2015}. The state of the art of such model involves the simulation of structure formation that includes the influence of baryonic matter (e.g. \cite{mocz19b,Veltmaat_2020}). The influence of FDM dynamics on the kinematics of baryons is relevant to establish bounds on the model, specially on the boson mass. Examples of the boson-baryon interaction are focused on the effects that FDM dynamics has on baryonic matter, which include the kinematic effects on clusters and black holes \cite{Bar-Or},  that influence the properties of structures like stellar disks \cite{Moczheating}, stellar streams \cite{AmoriscoLoeb}, as well as  the random motion due to the permanent interference of motion of FDM around galactic cores \cite{SchiveRandom}, even if these are dominated by a central soliton \cite{Dutta_Chowdhury_2021}.

Since these scenarios are studied with the most sophisticated numerical methods, the construction of spherically symmetric stationary configurations might seem too simple. Nevertheless, they are valuable, for example they can be used as test beds for numerical codes used within the FDM simulations, which are very sophisticated, and yet no standard comparison has been done. They can also be used as {\it ab initio} configurations with controlled initial virialization or initial total energy, that can help to numerically study the mergers of FDM+baryon structures.

These configurations may also become astrophysically interesting. Perhaps the most appealing property of these solutions is that, likewise Newtonian Boson Stars \cite{GuzmanUrena2004}, they behave as attractor solutions for at least a basin of attraction of parameters near equilibrium solutions. Solutions without the fluid where found to be attractors in simple scenarios before \cite{GuzmanUrena2006,BernalGuzman2006a} and later on they were found to be attractor solutions in average time in structure formation simulations \cite{Schive:2014dra,mocz19b}. The possibility that these solutions can be attractors within similar scenarios that now include baryonic matter fully coupled to the FDM could be very interesting to the FDM model.

Concerning the construction of solutions, the summary of assumptions is the following: spherical symmetry, harmonic time dependence of the wave function describing de bosons, the fluid obeys a polytropic Equation of State (EoS), the densities of bosons and the fluid source together a gravitational potential that obeys Poisson equation, and finally we consider that the configuration is isolated.

Stationary solutions were constructed for a finite but wide parameter space that covers the regimes when bosonic matter dominates over the fluid and the converse. These solutions are evolved to learn about their stability as response to perturbations due to numerical errors. The evolution is carried out using a code based on standard numerical methods summarized as follows: the Method of Lines is used for the evolution on a uniformly discretized numerical domain, we use finite difference stencils for the spatial operators in Schr\"odinger equation, the finite volume discretization is used to describe the evolution of the fluid along with second order variable reconstructors and the HLLE approximate Riemann solver, and finally Poisson equation is solved with a Multigrid elliptic solver \cite{AlvarezGuzman2022}.

Our findings are that when using the polytropic EoS during the evolution, that is, when the pressure is forced to obey this restriction, all the solutions in the parameter space are stable. Nevertheless when using the ideal gas EoS, the solutions are unstable, meaning that the system relaxes through expansion, which in turn cools the gas down until it approaches a less compact nearly stationary configuration.

The contents of the manuscript is as follows. In Section \ref{sec:model} we describe the system of equations that rule the dynamics of bosons and the fluid together. In Section \ref{sec:results} we show the equilibrium configurations constructed as well as the evolution of a representative sample of solutions. Finally in Section \ref{sec:conclusions} we draw some conclusions.

\section{Equations}
\label{sec:model}

The system is governed by the Schrodinger-Poisson-Euler (SPE) equations that we write as

\begin{eqnarray}
i\partial_t \Psi = -\frac{1}{2}\nabla^2\Psi + V \Psi, \label{eq: Schrodinger} \\
\partial_t \rho + \div\left(\rho\vec{v}\right)  =  0, \label{eq: mass conservation} \\
\partial_t \left(\rho\vec{v}\right) + \div\left(\rho\vec{v}\otimes\vec{v} + pI\right) = -\rho\grad V, \label{eq: momentum conservation} \\
\partial_t E + \div\left[\vec{v}(E+p)\right]  =  -\rho\vec{v}\cdot\grad V, \label{eq: energy conservation}  \\
\nabla^2 V  =  \rho + |\Psi|^2 \label{eq: poisson},  
\end{eqnarray}

\noindent where the dynamics of the FDM is described by the order parameter $\Psi$ which represents the state of an ensemble of bosons with mass density $|\Psi|^2$, while the dynamics of the fluid is described by Euler equations where each volume element has mass density $\rho$, velocity $\vec{v}$, pressure $p$, total energy $E = \rho\left(e+\frac{1}{2}|\vec{v}|^2\right)$, internal specific energy $e$. Finally, boson and fluid densities source the gravitational potential $V$. These equations are written in such a way that the constants $m_B$, $\hbar$ and $4\pi G$ are absorbed by redefinitions of all variables. The dictionary between code variables in system (\ref{eq: Schrodinger})-(\ref{eq: poisson}) and physical quantities identified with tilde, is explicitly as follows: $\Psi=\frac{\sqrt{G m_B^3} R_0^2}{\hbar}\tilde{\Psi}$, $\rho = \frac{G m_{B}^2 R_0^4}{\hbar^2} \tilde{\rho}$, $\vec{v} = \frac{m_B R_0}{\hbar} \tilde{\vec{v}}$, $e = \left(\frac{m_B R_0}{\hbar}\right)^2 \tilde{e}$, $p = \frac{G m_B^4 R_0^6}{\hbar^4}\tilde{p}$, $E = \frac{G m_B^4 R_0^6}{\hbar^4} \tilde{E}$, $V=\left(\frac{m_{B}R_0}{\hbar}\right)^2\tilde{V}$, with spatial coordinates $x_i=\frac{\tilde{x_i}}{R_0}$ and time $t=\frac{\hbar}{m_B R_0^2}\tilde{t}$, where $R_0 = \frac{\hbar^2}{G m_{B}^2 M_0}$ is the relation between given mass $M_0$ and length  $R_0$ scales. Notice that the system (\ref{eq: Schrodinger})-(\ref{eq: poisson}) is indeterminate since it is defined by seven equations for eight variables, that we close as usual, with an equation of state (EoS) $p = p(\rho, e)$.\\

A useful property of the SP subsystem is its invariance under rescaling by means of a factor $\lambda$ \cite{Ruffini:1969,GuzmanUrena2004,Mocz:2017wlg}:

\begin{equation}
\{t,\vec{x},V,\Psi\} \to \{\lambda^{-2}t, \lambda^{-1}\vec{x},\lambda^2 V, \lambda^2\Psi\},
\label{eq:SP scaling}
\end{equation}

\noindent that helps rescaling the bosonic part of the system. For the complete SPE system without EoS to be invariant before this rescaling, it is necessary to rescale the hydrodynamic variables as follows

\begin{equation}
\{\rho, \vec{v}, E, p, e \} \to \{\lambda^4 \rho, \lambda^{-1}\vec{v}, \lambda^2 E, \lambda^2 p, \lambda^{-2} e\}.
\label{eq:Euler scaling}
\end{equation}

\noindent The set of equations (\ref{eq: Schrodinger})-(\ref{eq: poisson}) can be viewed as an IVP and solved for some initial conditions on all the variables. A number of scenarios were studied in \cite{AlvarezGuzman2022} that illustrate the dynamics of the bosonic and fluid components.

Particularly interesting configurations are equilibrium solutions, for example for the fluid alone it is possible to construct Newtonian versions of Tolman-Oppenheimer-Volkoff equilibrium configurations \cite{AlvarezGuzman2022}, whereas for the purely bosonic components it is possible to construct Newtonian Boson Stars \cite{GuzmanUrena2004}. These configurations serve as basic models of not so compact TOV stars and Newtonian Bosonic Stars respectively.

\subsection{Stationary solutions}

In this paper we study the construction of stationary configurations of the system (\ref{eq: Schrodinger})-(\ref{eq: poisson}) for the two components at the same time and study their basic properties. These solutions are constructed by assuming spherical symmetry and stationarity. If the variables are described with spherical coordinates, the wave function can be written as $\Psi(r,t) = \psi(r) e^{i\omega t}$ where $\psi$ is a real function, which guarantees that $|\Psi|^2$ is time-independent. Time independence also implies that the fluid is in hydrostatic equilibrium, which implies that $\rho=\rho(r)$, $p=p(r)$, $\vec{v}=\vec{0}$. Time independence of $|\Psi|^2$ and $\rho$ imply that the gravitational potential is time-independent too $V=V(r)$. With these considerations the SPE system reduces to the following set of ODEs along the radial coordinate $r$:

\begin{eqnarray}
-\dfrac{1}{2r^2}\dfrac{d}{dr}\left(r^2\dfrac{d\psi}{dr} \right) + V\psi = -\omega \psi,\label{eq:Statpsi} \\
\dfrac{dp}{dr} = -\rho \dfrac{dV}{dr}, \label{eq:Statp}\\
\dfrac{1}{r^2}\dfrac{d}{dr}\left(r^2 \dfrac{d V}{dr}\right) = \rho + \psi^2,\label{eq:StatV}
\end{eqnarray}

\noindent which is closed using a polytropic equation of state $p = K\rho^{1+1/n}$, where $K$ is the polytropic constant and $n$ is the polytropic index. In general, this system of ODEs defines an eigenvalue problem for the constant $\omega$, which is solved numerically by means of the shooting method that uses the fourth-order Runge-Kutta method. For the numerical solution, we define the auxiliary variables $\phi:= r^2 \frac{d\psi} {dr}$ and $M := r^2\frac{dV}{dr}$, that reduce the system (\ref{eq:Statpsi})-(\ref{eq:StatV}) to the first order system

\begin{eqnarray}
\dfrac{d\psi}{dr} = \dfrac{\phi}{r^2}, \label{eq:psir}\\
\dfrac{d\phi}{dr} = 2\left(\omega + V\right)\psi r^2, \label{eq:phir} \\
\dfrac{dp}{dr} = -\rho \dfrac{M}{r^2}, \label{eq:pr} \\
\dfrac{dV}{dr} = \dfrac{M}{r^2} \label{eq:Vr}, \\
\dfrac{dM}{dr} = \left(\rho + \psi^2\right)r^2 \label{eq:Mr}, \\
p = K\rho^{1+1/n} \label{eq:poly}.
\end{eqnarray}

\noindent The type of solutions we are interested in are regular at the origin, localized, which requires the wave function as well as density and pressure to vanish as $r\rightarrow \infty$. Thus, appropriate boundary conditions are the following:

\begin{eqnarray}
\psi(0) & = & \psi_c, \\
p(0) & = & K\rho_c^{1+1/n}, \\
\phi(0) = M(0)  & = & 0, \\
\lim_{r\to\infty}\psi(r) =  \lim_{r\to\infty}V(r) & = & 0.
\end{eqnarray}

\noindent  Once the polytropic constant $K$ and polytropic index $n$ are fixed, it is possible to construct a family of solutions parametrized by the central values of the wave function $\psi_c$ and fluid density $\rho_c$. 

A comment is in turn. When the  two componentes are considered separately, the construction of a single stationary solution with $\psi_c=1$ suffices to produce the hole family of solutions of the Schr\"odinger-Poisson system using the scaling transformation (\ref{eq:SP scaling}). On the other hand, solutions of Euler-Poisson equations can be rewritten as the Lane-Emden equation with a change of variable $\rho = \rho_c \theta^n$ and thus obtain all the solutions from the solution with the condition $\theta (0) =1$ (e.g. \cite{AlvarezGuzman2022}). 

Unfortunately, for the two components coupled through the SPE system of equations, a scaling relation requires the polytropic constant to rescale like $K \to \lambda^{-2-4/n} K$, which is sufficient to prevent the construction of a single solution and then use the scaling relation to produce the whole family of solutions for a given pair $(K,n)$. The system (\ref{eq:psir}-\ref{eq:poly}) must be solved for each tuple of values $(\psi_c, \rho_c)$ and the invariance property can only  be used to construct new solutions with different values of the polytropic constant $K$. 

\subsection{Diagnostics of the solutions}

Two important quantities in the stationary case are the masses $m_{\rho}$ and $m_{\psi}$ of each component as function of the radial coordinate $r$, calculated with the expressions

\begin{equation}
m_{\rho}(r) := 4\pi \int_0^r \rho(r') (r^{\prime})^2 dr^{\prime}, 
\label{eq:mgas}
\end{equation}
\begin{equation}
m_{\psi}(r) := 4\pi \int_0^r \psi(r')^2 (r^{\prime})^2 dr^{\prime}.
\label{eq:mwave}
\end{equation}

\noindent With these integrals it is possible to define the total mass of each component as $M_\psi = \lim_{r\to\infty}m_\psi(r)$ and $M_{\rho} = m_\rho(R_\rho)$ where $R_\rho$  is the first zero of the density $\rho$. We use these two values to define the mass ratio between the fluid and the bosons 

\begin{equation}
MR = M_\rho / M_\psi\label{eq:mr},
\end{equation} 

\noindent that in turn determines the dominance of each material component.

Another useful quantity is the compactness of the solutions $M_{\rho}/r_{\rho,95}$ and $M_{\psi}/r_{\psi,95}$ where the radii $r_{\rho,95}$ and $r_{\psi,95}$ define the sphere containing $95\%$ of the masses $M_{\rho}$ and $M_{\psi}$ respectively.

Also interesting are the rotation curves for test particles traveling around the configuration in circular orbits is interesting to model galaxies made of FDM. The circular velocity is given in terms of the mass functions above:

\begin{equation}
v(r) = \sqrt{\dfrac{m_{\rho}(r) + m_{\psi}(r)}{4\pi r}}.
\label{eq:vRT}
\end{equation}

\noindent In the case of bosonic matter alone, stationary solutions are considered to be galactic cores of dark matter, that serve to fit rotation curves of dwarf galaxies \cite{Bernal_2017} and the expression above would help to measure the effects of the addition of the fluid on the RC.

\section{Results}
\label{sec:results}

\subsection{Equilibrium Configurations}
\label{subsec: CEC}

\begin{figure}
\includegraphics[width=8cm]{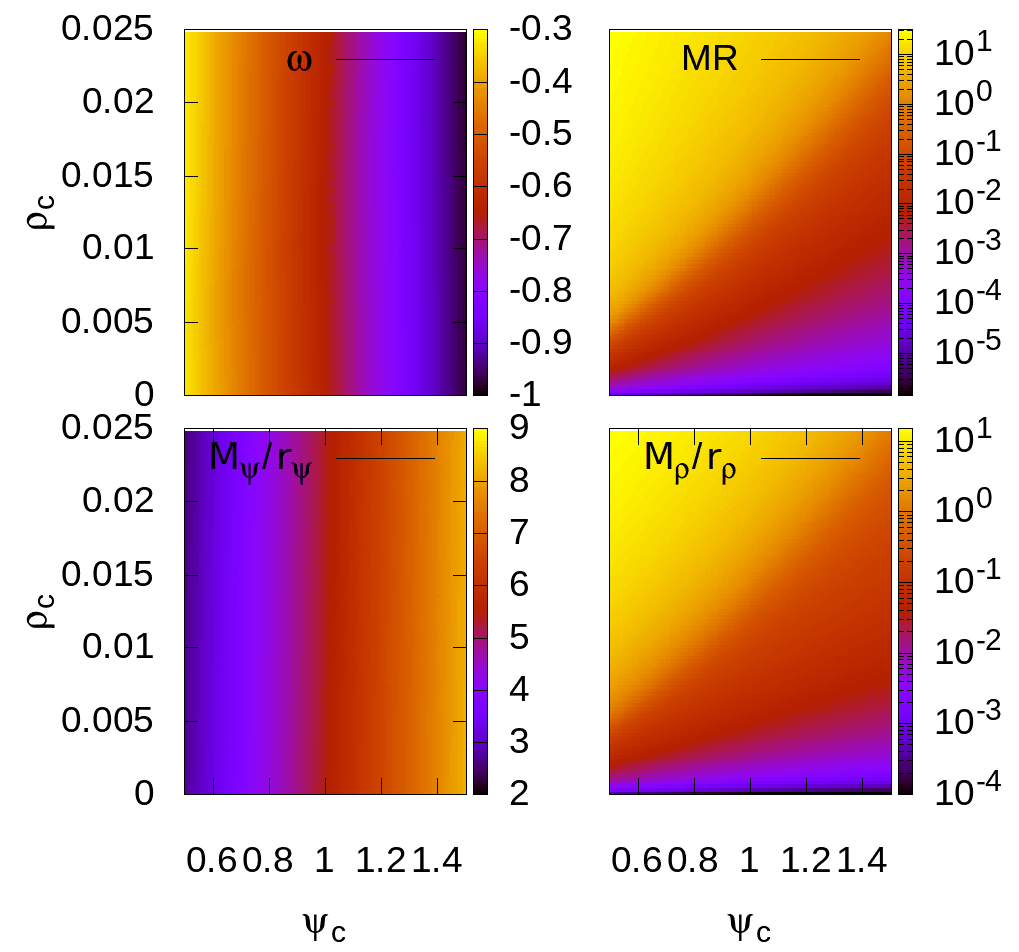}
\caption{In the upper part we show the oscillation frequency $\omega$ and $MR$ as function of the initial parameters $(\psi_c,\rho_c)$.At the bottom the compactness of the wave function $M_\psi / r_{\psi,95}$ and that of the polytrope $M_\rho / r_{\rho,95}$.}
\label{fig:equilibrium}
\end{figure}

We explore the parameter space $(\psi_c, \rho_c)\in[0.5,1.5]\times[0,2.5\times10^{-2}]$ for a polytropic constant $K=10$ and a polytropic index $n=3$. Consider that for a given value of $K$, a whole family of solutions with the scaling $K \to \lambda^{-2-4/n} K$. The polytropic index is set to $n=3$ because it corresponds to an adiabatic index $\gamma=5/3$, suitable for an ideal monoatomic gas. In Figure \ref{fig:equilibrium} we show the eigenfrequency $\omega$ as function of the parameters $(\psi_c,\rho_c)$, where we notice that the frequency is a growing function of $\psi_c$ and $\rho_c$. We also  show the mass ratio $MR$, which reaches its maximum in the regions where $\psi_c$ has its minimum value and $\rho_c$ its maximum within the parameter space explored, plus it is an increasing function on the variables $(\psi_c,\rho_c)$ so that a solution can be uniquely determined from this value. At the bottom we show the compactness of both, the wave function and the polytrope. Observe that the compactness of the wave function depends slightly on the central density $\rho_c$. The compactness of the polytrope for the smaller $\psi_c$ and the larger $\rho_c$ correspond to a bigger compactness, in the same way as it happens with the mass ratio $MR$. Notice that the value of $MR$ covers a wide range of bosonic or ferminonic matter dominance, ranging from $10^{-5}$ to 10. If these solutions are to be a toy  galactic model, it would cover a considerable variety of dark to luminous matter contributions in a galaxy, from dwarfs dominated by dark matter to those dominated by luminious matter.

\begin{figure}
\includegraphics[width=4cm]{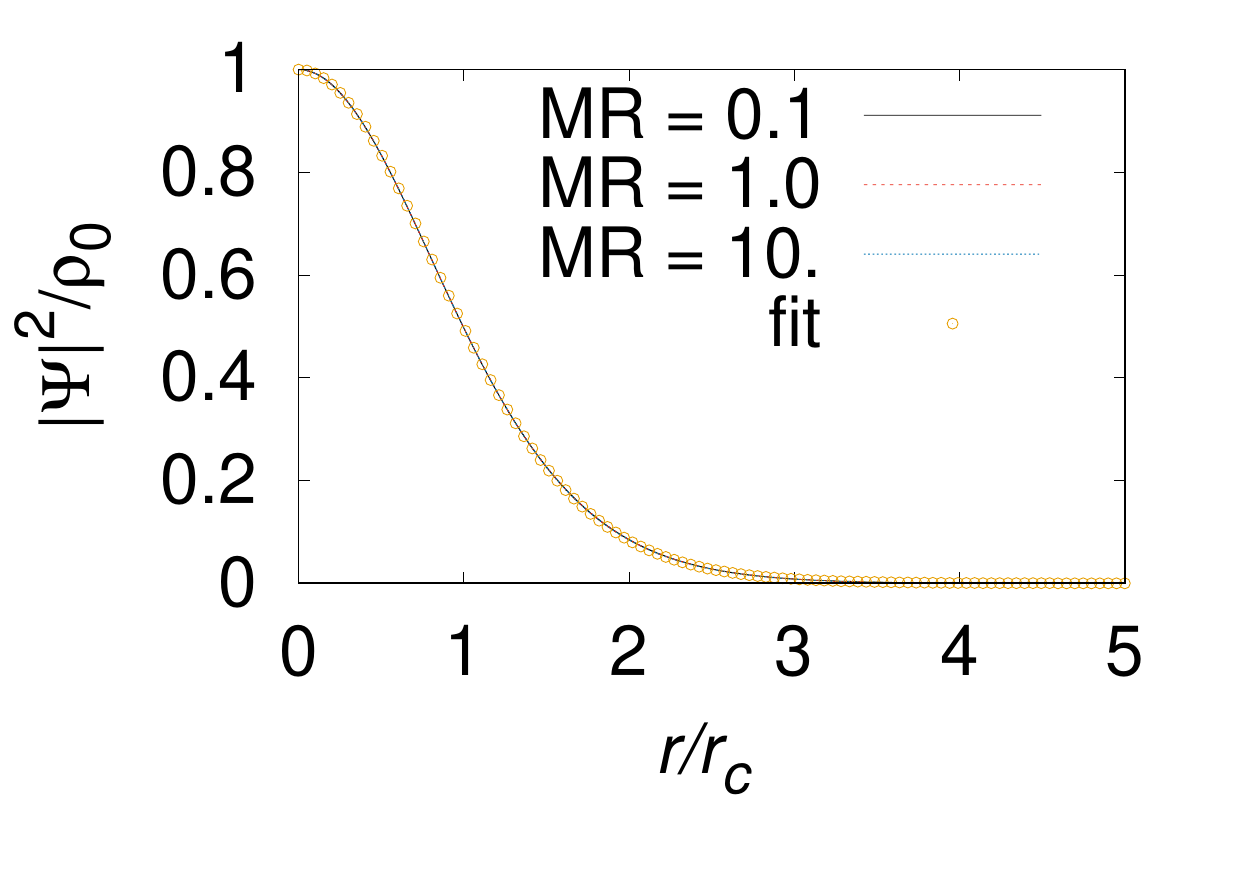}
\includegraphics[width=4cm]{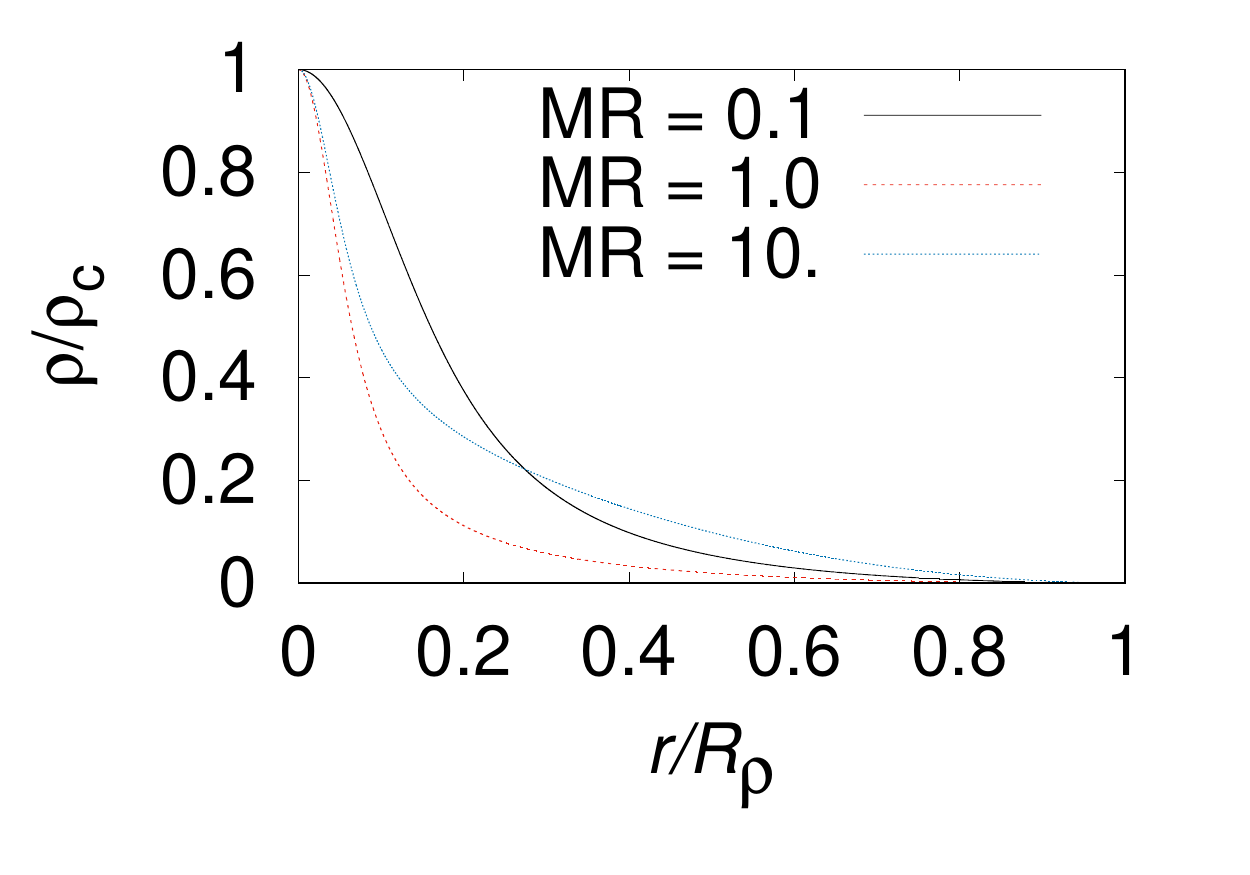}
\includegraphics[width=4cm]{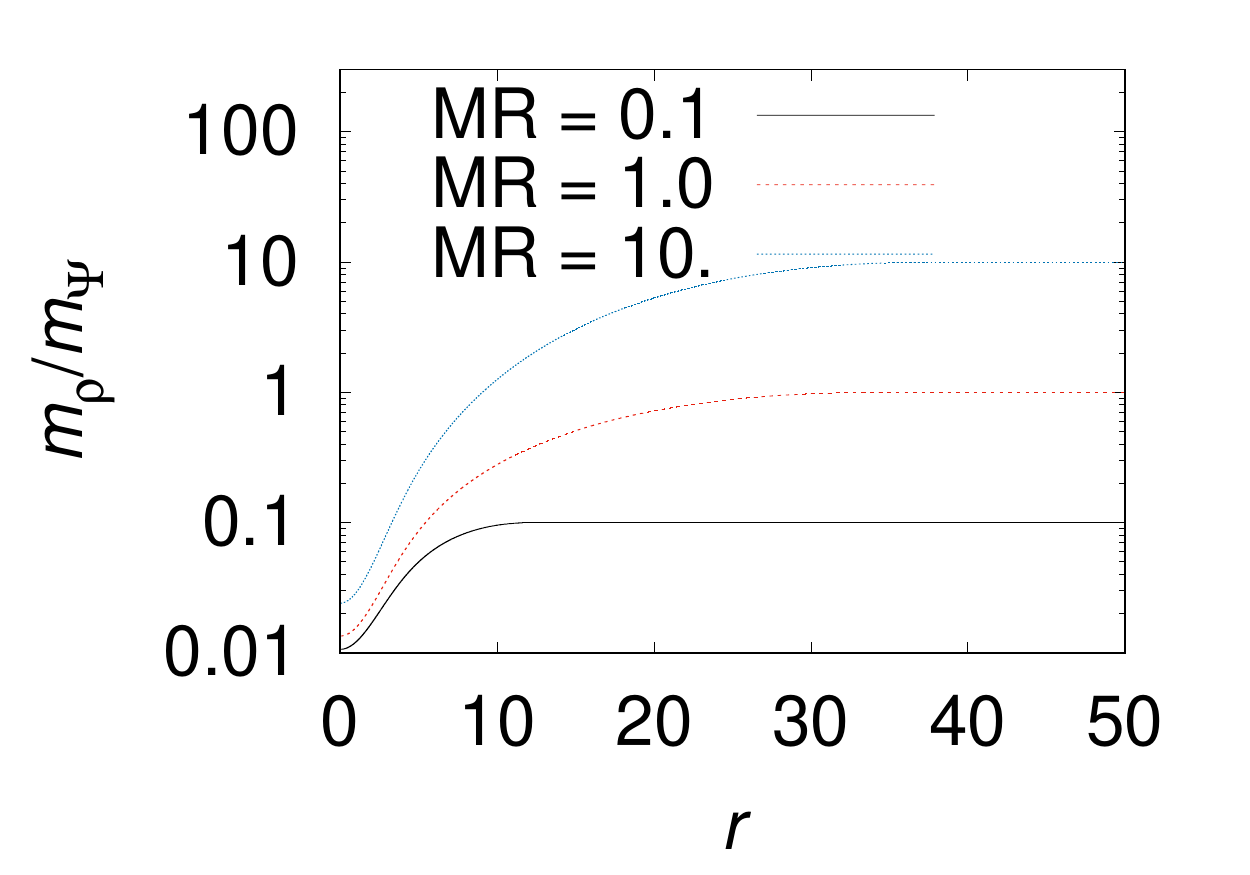}
\includegraphics[width=4cm]{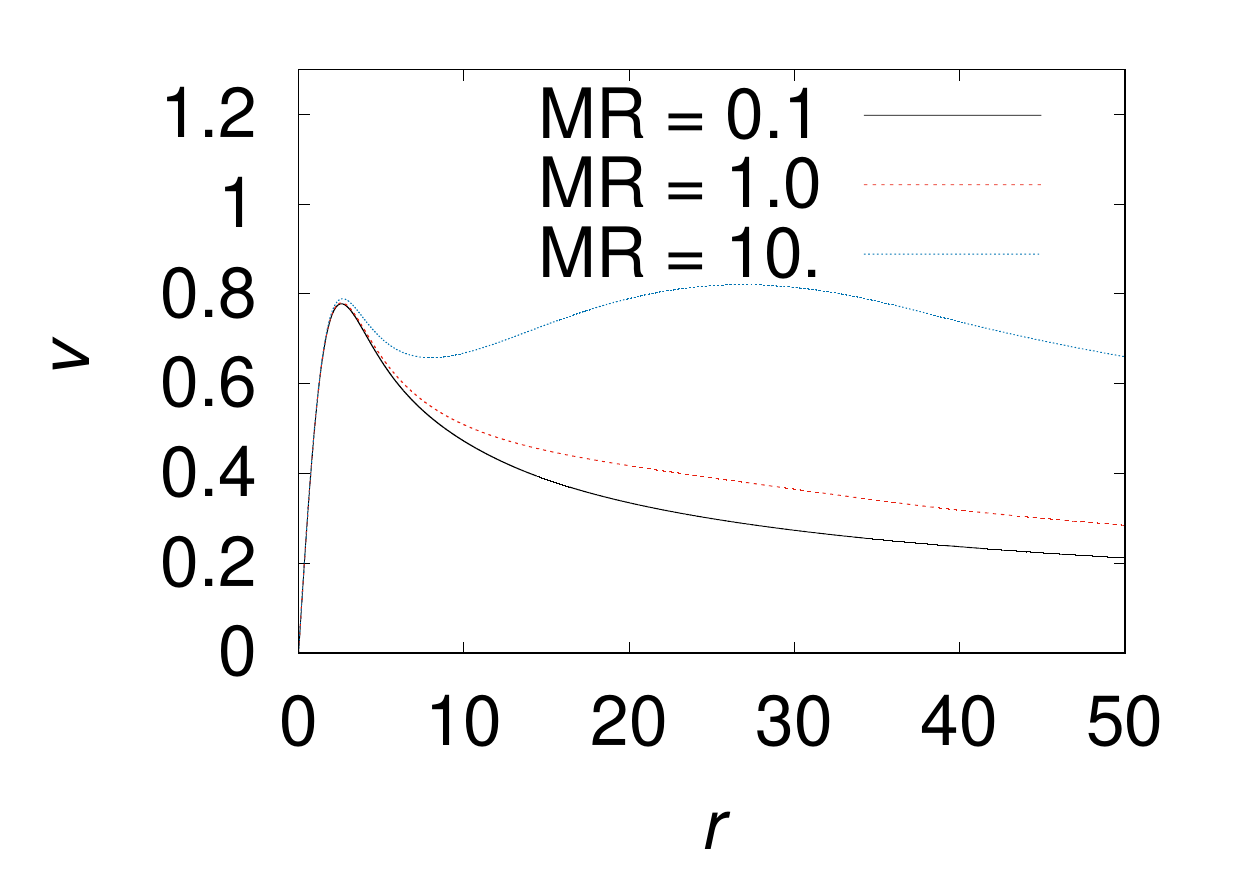}
\caption{The top-left side shows the density of the wave function as a function of the normalized radial coordinate with respect to the radius of the core $r_c$, which is an adjustment parameter given in Eq. (\ref{eq: empirical soliton}). On the top-right side, the density of the gas as a function of the normalized radial coordinate with respect to the radius $R_{\rho}$. In the bottom-left $m_{\rho}/m_{\psi}$ as a function of the radial coordinate in which we can see that the asymptotic value tends to the corresponding values of the mass ratio $MR=0.1, 1$ and 10. In the bottom-right the rotation curve of these three solutions as a function of the radial coordinate.}
\label{fig:density equilibrium}
\end{figure}

To show in more detail the radial behavior of the solutions we choose three equilibrium solutions of this parameter space for the central values $\psi_c = 1$ and $\rho_c \approx 1.06\times 10^{-2},1.34\times 10^{-2}$ and $2.41\times 10^{-2}$,    shown in Figure \ref{fig:density equilibrium}. On the top-left we have the wave function density and on the top-right the density of the gas, which satisfy the mass ratio $MR=0.1, 1$ and $10$. Likewise in the purely bosonic case, the density $|\Psi|^2$ of the wave function can be fitted with the solitonic empirical profile:

\begin{equation}
|\Psi|^2=\rho_{solitonic}(r) = \rho_{0}\left[1+0.091 \left(\dfrac{r}{r_c}\right)^2\right]^{-8},
\label{eq: empirical soliton}
\end{equation}

\noindent where $\rho_0 = \rho_0(\rho_c,r_c)$ and $r_c = r_c(\rho_c)$ are functions of the central density $\rho_c$ which in the three representative cases has values $\rho_0=1$ and $r_c \approx 1.30, 1.298, 1.291$, which also indicates that when the value of the fluid central density increases, the density of the wave function tends to be more compact. The radius $r_{soliton} \approx 3.5 r_c$ is the one that contains most of the configuration mass \cite{Mocz:2017wlg}, in our cases $r_{soliton} \approx 4.55,4.54,4.52$ and the radius of the gas where the density vanishes for these three solutions is $R_\rho \approx 12.7,35.7,39.1$ respectively. Then we can see that in these representative cases, the gas distribution extends out of the solitonic core. At the bottom-left side we show the quantity $m_{\rho}/m_{\psi}$ as function of the radius, in which one notices that these quantities satisfy the asymptotic relation of mass $MR=0.1,1$ and $10$. At the bottom-right the rotation curve of these solutions in which we can see that increasing the mass ratio produces an increase in the tangent speed of a test particle.

\subsection{Stability of equilibrium solutions}

An essential property of solutions is stability, that we evaluate in two ways. The {\it first method} consists in the evolution of equilibrium configurations by means of the  polytropic EoS used to construct the solutions, we recall that a process where the constant $K$ remains constant everywhere is equivalent to the entropy being uniform, and the evolution is carried out assuming that the process will always be isentropic. The {\it second method} assumes that at initial time the process is not only isentropic but also adiabatic, in this way we can identify the polytropic index $n$ with the adiabatic index $\gamma = 1+1/n$ of an ideal gas with EoS $p = (\gamma - 1)\rho e$ where at initial time $e= K \rho^{\gamma-1} / (\gamma - 1)$, then the evolution can be non-isentropic. 

\begin{figure}
\includegraphics[width=4cm]{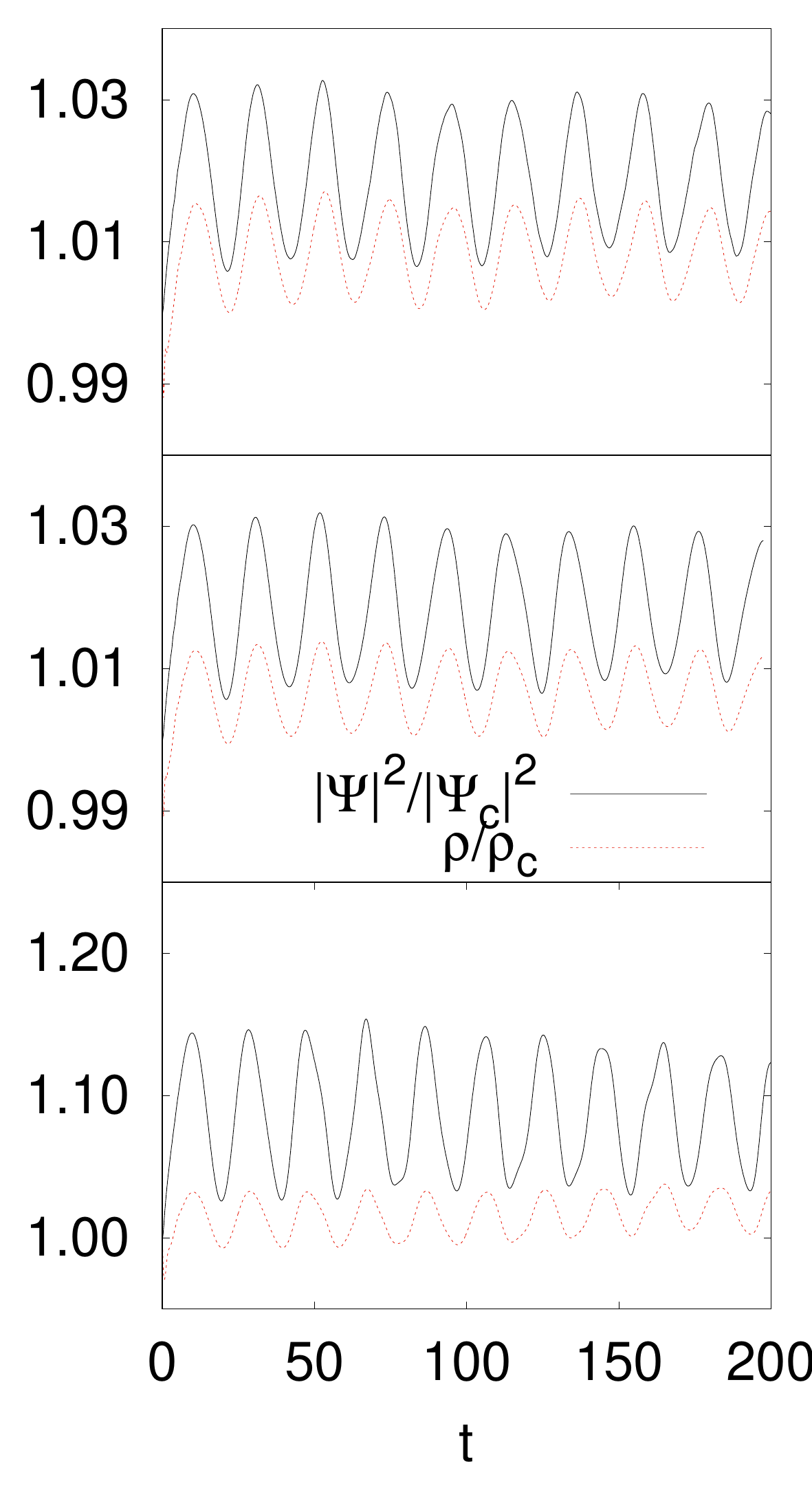}
\includegraphics[width=4cm]{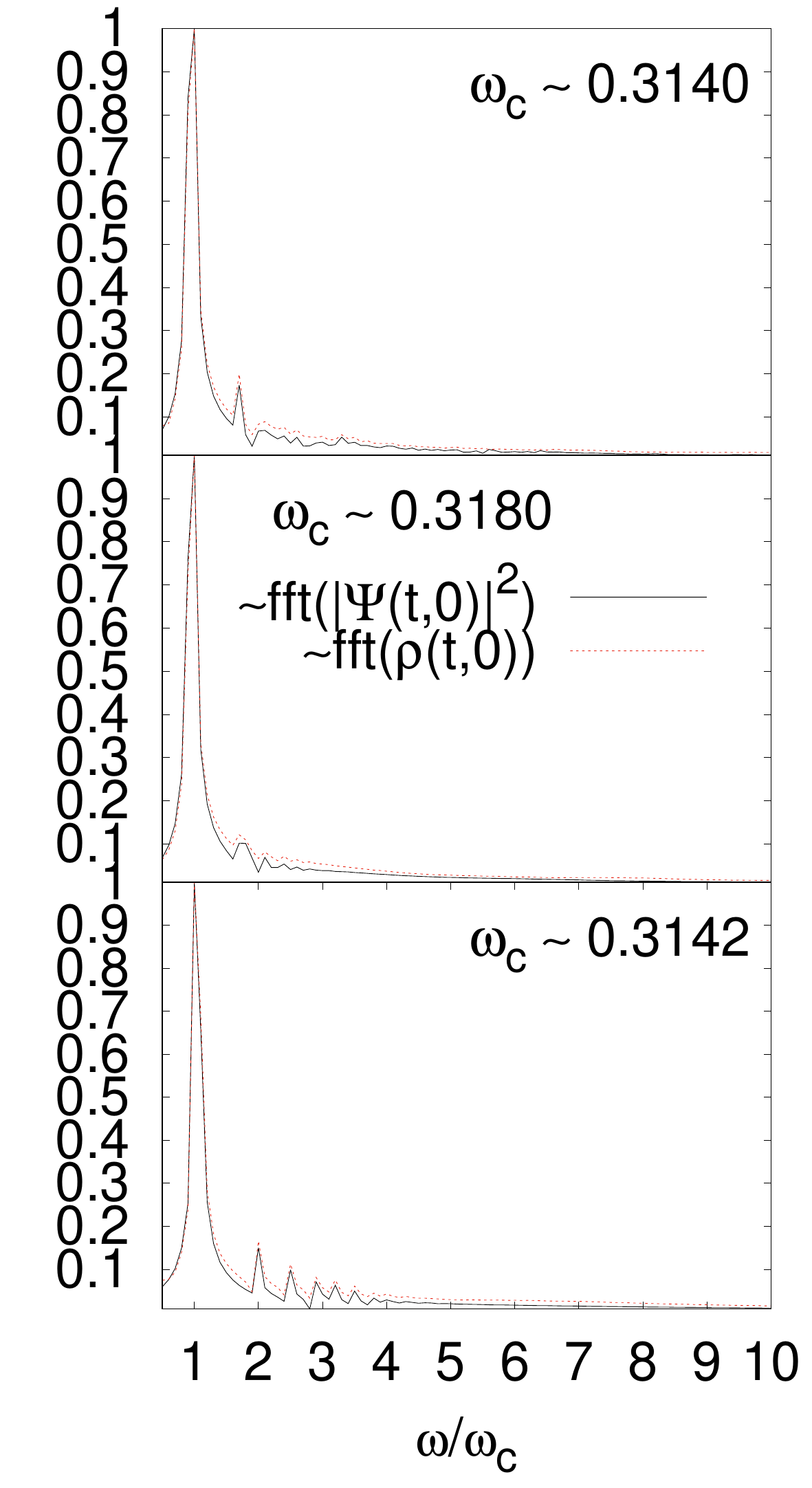}
\caption{Evolution using the polytropic EoS for the fluid. At the left we show the evolution of the central densities for the polytrope $\rho/\rho_c$ and wave function $|\Psi|^2/|\Psi_c|^2$. At the right side we show the Fourier Transform of the central densities. From top to bottom the cases with $ MR=0.1$, $MR=1$ and $MR=10$, whose oscillation modes have peak frequencies $\omega_c \sim 0.3140, 0.3180$ and $0.3142$ respectively.}
\label{fig:polyrhomax}
\end{figure}

\begin{figure}
\includegraphics[width=4cm]{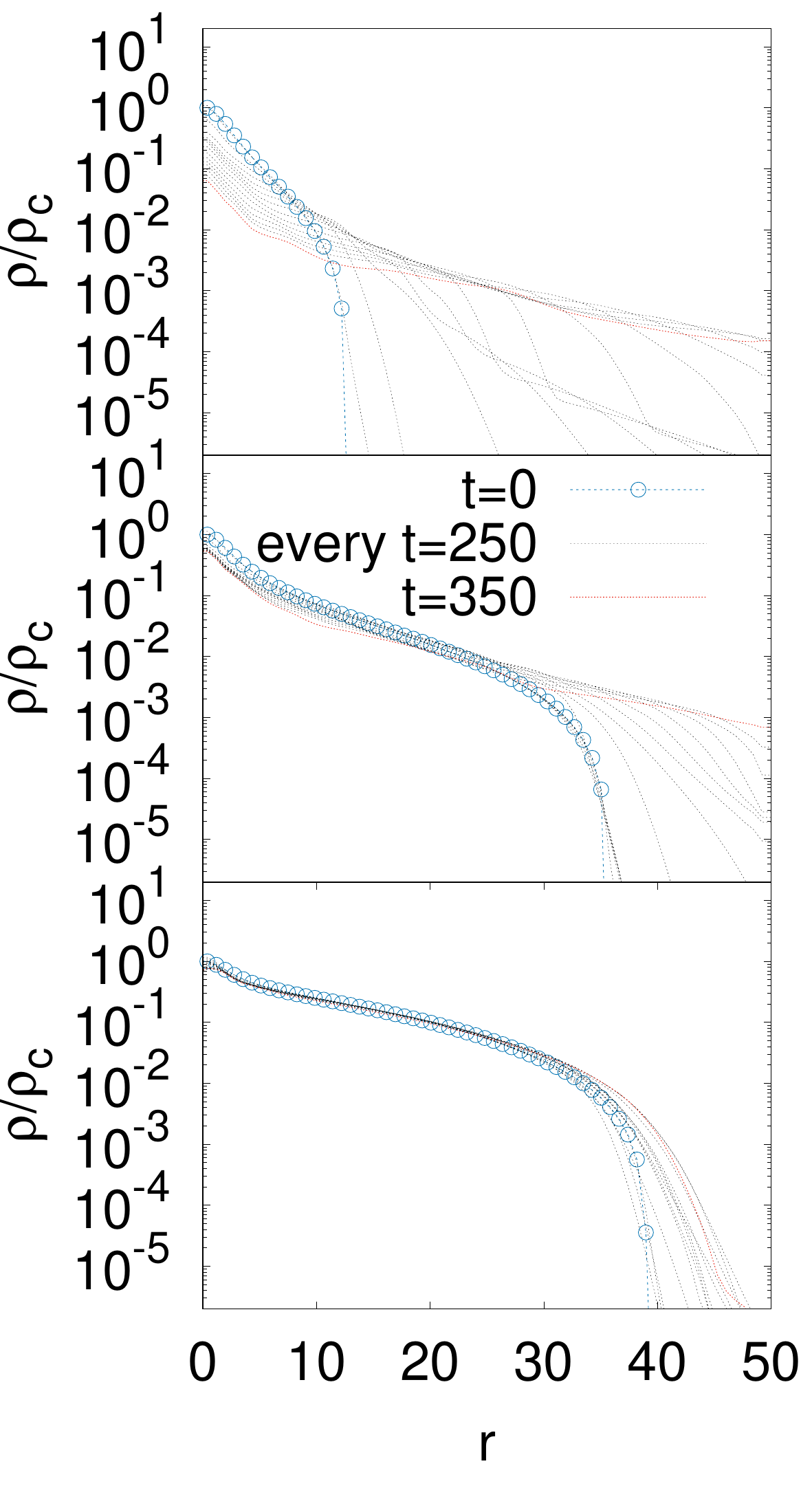}
\includegraphics[width=4cm]{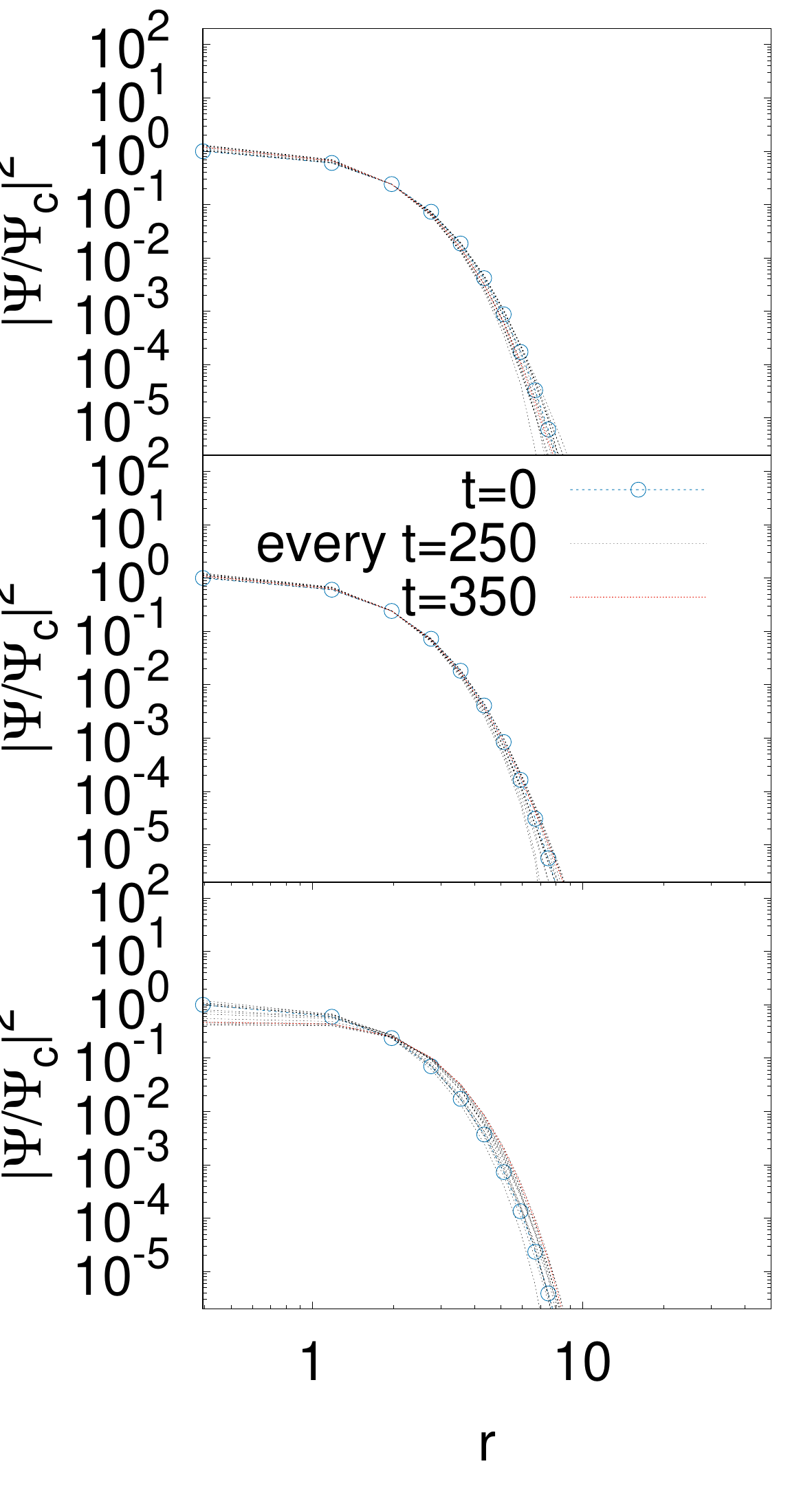}
\caption{Snapshots of the gas density and the wave function density for evolutions of the three representative configurations with $MR=0.1$, 1 and 10 from top to bottom. It can be seen that the density of the wave function remains oscillating near the initial configuration while the gas tends to redistribute to a new configuration.}
\label{fig:snapsEoSideal}
\end{figure}

For the two methods we choose the sample of three representative solutions with tuples $(\psi_c=1,\rho_c)$, where the central fluid density $\rho_c$ is chosen such that the mass ratio of the configuration has the values $MR=0.1,1$ and $10$, in Eq. (\ref{eq:mr}). These solutions are interpolated into a three-dimensional cubic domain $ D = [-50,50]^3$ described with Cartesian coordinates, and used as initial conditions to start the evolution of the system of equations (\ref{eq: Schrodinger})-(\ref{eq: poisson}) in order to evaluate the stability properties.

Numerical specifications for the evolution are the spatial resolution of $D$, which is discretized with a Fixed Mesh Refined domain (see e.g. \cite{Grupo2014}), with a base resolution $\Delta x = \Delta y = \Delta z =1/3$ and time resolution $\Delta t= 0.05 \Delta x^2$, using two refinement boxes constructed doubling resolution and centered at the coordinate origin. 

{\it Numerical methods for the evolution.} The equations are solved using the Method of Lines with a third order TVD Runge-Kutta integrator. The right hand side of Schr\"odinger equation (\ref{eq: Schrodinger}) is discretized using fourth order accurate Finite Difference stencils. The right hand sides of hydrodynamics (\ref{eq: mass conservation},\ref{eq: momentum conservation},\ref{eq: energy conservation}) are discretized using a Finite Volume approach, along with second order variable reconstructors and the HLLE approximate flux formula \cite{Toro2009,AlvarezGuzman2022}. Finally, Poisson equation (\ref{eq: poisson}) is solved at each intermediate step of the Runge-Kutta integrator for $V$ using a Multigrid algorithm \cite{multigrid}.

{\it Evolution with the polytropic EoS.} The central density of both matter distributions as functions of time, normalized with their initial value is shown in the first column of Figure \ref{fig:polyrhomax}. The second column shows the Fourier Transform of the time series, with peak frequencies $\omega_c = 0.3140, 0.3180$ and $0.3142$ for the cases with $MR=0.1,1$ and $10$ respectively. It can be observed that the solutions oscillate near the equilibrium solution, which indicates that the solutions are stable. Time series and their Fourier Transform also indicate the correlation between phase and frequency of oscillations for the two components.

{\it Evolution using the ideal gas EoS.} Figure \ref{fig:snapsEoSideal} shows snapshots of de fluid and wave function densities for the three representative cases with $MR=0.1$, 1 and 10 from top to bottom. It is observed that when promoting an equilibrium solution with an EoS of ideal gas, the gas density distribution clearly departs from the equilibrium solution, whereas the bosonic density is less affected. The density of the gas tends to redistribute to a new stable solution faster for the higher value of $MR$.

\begin{figure}
\includegraphics[width=4cm]{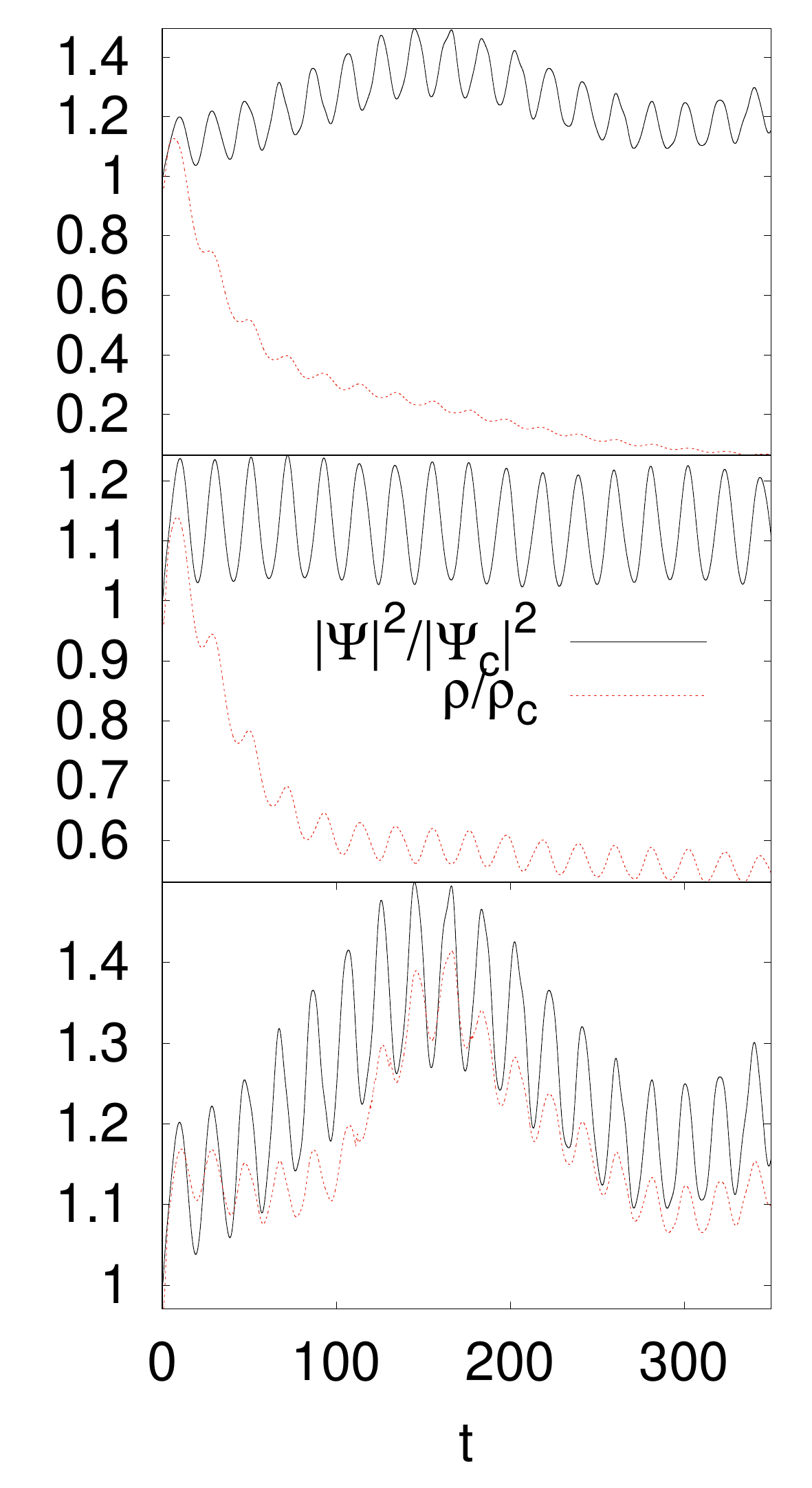}
\includegraphics[width=4cm]{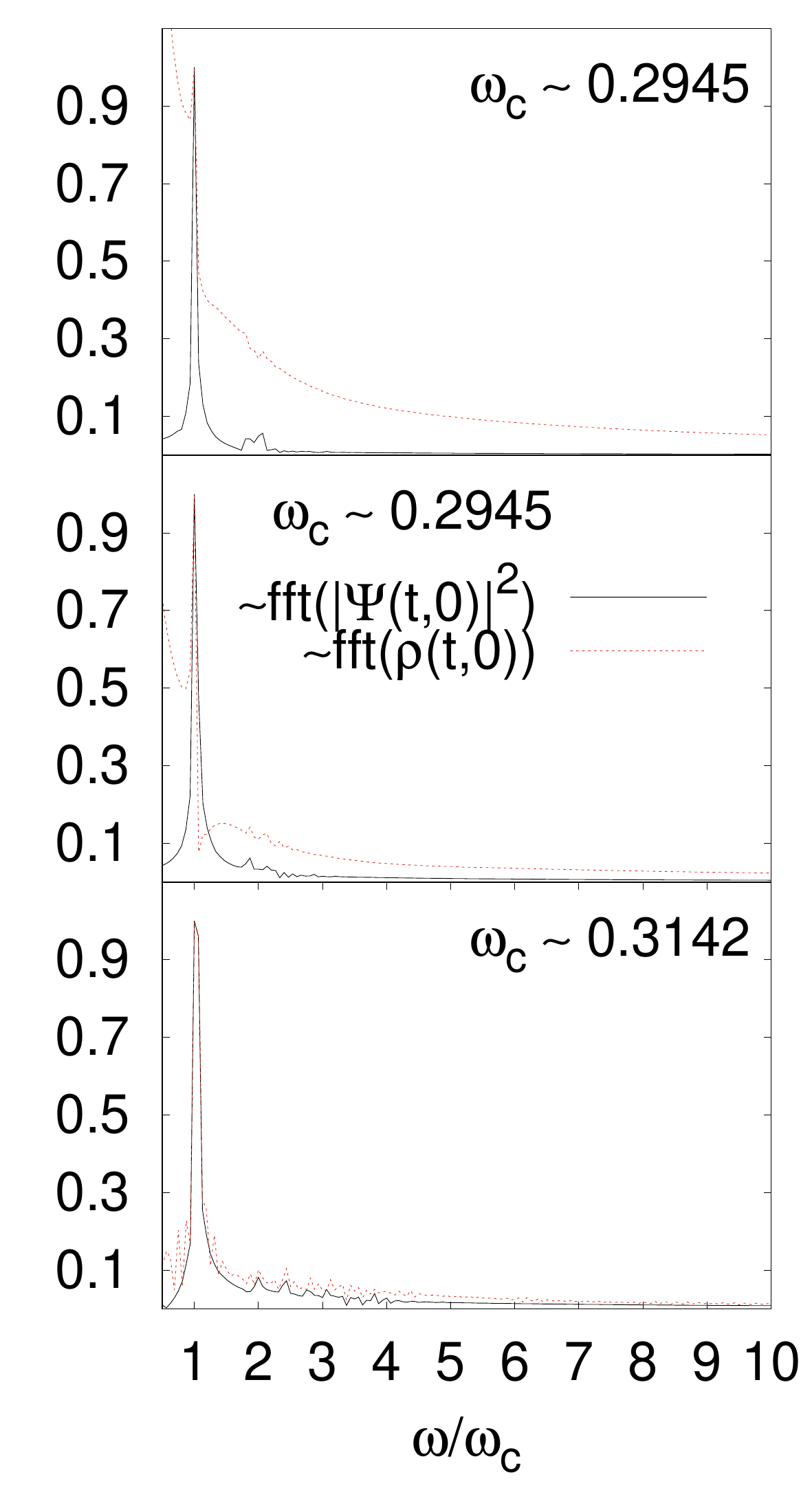}
\caption{Evolution using the ideal gas EoS for the fluid. At the left we show the evolution of the central densities of the fluid $\rho/\rho_c$ and wave function $|\Psi|^2/|\Psi_c|^2$. At the right we show the Fourier Transform of the central densities. From top to bottom we show the cases with $ MR=0.1,1,10$, whose oscillation modes have peak frequencies $\omega_c \sim 0.2945$ for the first two cases and $\omega_c \sim 0.3142$ for the last one.}
\label{fig:idealrhomax}
\end{figure}

In Figure \ref{fig:idealrhomax} we show the time series of central densities of the two components for $MR=0.1,1,10$. In the case $MR=0.1$ the fluid density decreases considerable, whereas for $MR=10$ the fluid density is less disturbed, while the wave function density is more disturbed in this case due to the interaction with the gas. In all three cases the Fourier Transform of the time series shows a correlation between the oscillations of the two components.

\begin{figure}
\includegraphics[width=6cm]{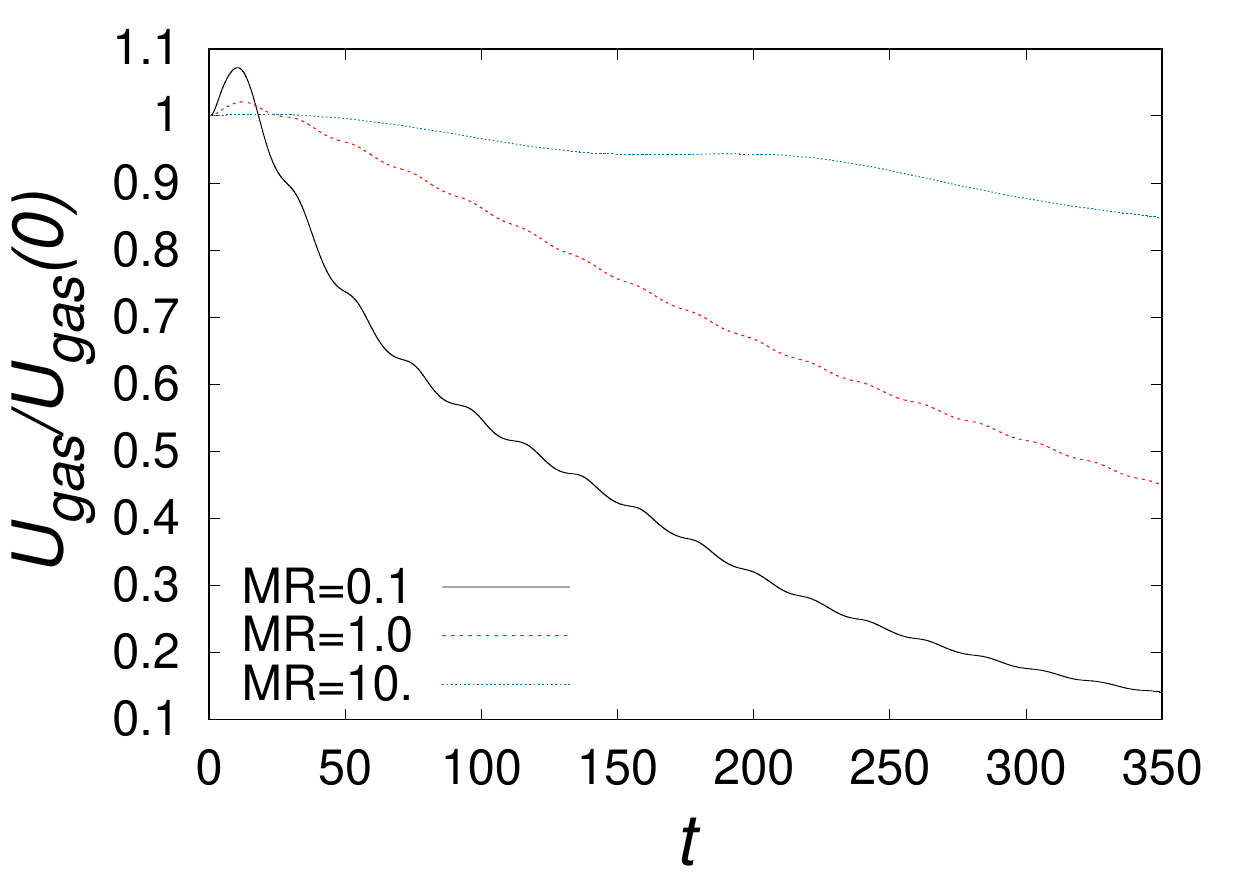}
\caption{Total internal energy of the ideal gas as function of time for the initial conditions corresponding to the values of the mass ratio $MR=0.1,1$ and 10. The internal energy has an initial growing transient corresponding to a compression and increase of temperature, and later on it decreases during the relaxation and cooling process.}
\label{fig:idealenergy}
\end{figure}

The bosonic distribution remains oscillating nearly around the equilibrium density profile, while the fluid does not. The ideal gas  was initially in an isentropic and adiabatic process, however gravitational energy dominates over the internal energy of the gas, and the volume element of the gas begins to decrease, which in turn increases the temperature. Once the maximum compression is achieved, the internal energy starts dominating over the gravitational energy, which triggers an expansion process that cools the gas down as can be seen in Figure \ref{fig:idealenergy}, where the total internal energy is obtained from the expression $U_{gas} := \int_D \rho e d^3x$.

A comment related to the evolution of these solutions is in turn. The instability of the solutions evolved assuming the fluid obeys an ideal gas EoS y due to the non-adiabatic nature of the process, specifically a compression followed by an expansion. This has to be contrasted with the instabilities with expansion in the general relativistic case for Boson Stars illustrated in \cite{Guzman2004BS,Guzman2009}, which is due to the binding energy that corresponds to a gravitationally unbound state.

\section{Conclusions}
\label{sec:conclusions}

We constructed stationary spherically symmetric solutions of the Scr\"odinger-Euler-Poisson system, parametrized by the central density of the fluid and the central value of the wave function. This was done for a wide range of parameter combinations represented mainly by the mass ratio between the fluid and the bosonic masses.\\

We also tested the stability of these solutions by evolving the solutions in time with a three-dimensional code that added perturbations beyond spherical symmetry due to truncation errors. Our findings are that all solutions when enforced to obey the polytropic EoS are stable, whereas when they are evolved using an Ideal Gas EoS all of them are unstable and relax towards a configuration with different mass distributions.

Even though these solutions and their stability are worth by themselves, we expect these solutions serve as test-beds for codes that simulate Fuzzy Dark Matter with baryons. Another potential application could be that these solutions serve as basic toy models of FDM+baryon configurations in the simulation of mergers that may give rise to a wide variety of galaxies within this dark matter model.



\section*{Acknowledgments}
Iv\'an \'Alvarez receives support within the CONACyT graduate scholarship program under the CVU 967478.
This research is supported by grants CIC-UMSNH-4.9 and CONACyT Ciencias de Frontera Grant No. Sinergias/304001. The runs were carried out in the facilities of the IFM-UMSNH.


\bibliography{BECDM}

\end{document}